\documentclass[12pt,fleqn]{iopart}

\usepackage{graphicx}
\usepackage{float}
\usepackage{placeins}

\begin{document}

\title{Propagation in 3D of microwaves through density perturbations}

\author{T R N Williams$^{1,2}$, A K\"{o}hn$^{3}$, M R O'Brien$^{2}$ and R G L Vann$^{1}$}

\address{$^{1}$York Plasma Institute, Department of Physics, University of York, York, YO10 5DD, UK}

\address{$^{2}$EURATOM/CCFE Fusion Association, Abingdon, OX14 3DB, UK}

\address{$^{3}$IGVP, Universit\"{a}t Stuttgart, Pfaffenwaldring 31, D-70569 Stuttgart, Germany}

\ead{trnw500@york.ac.uk}

\begin{abstract}
Simulations using 3D and 2D full-wave codes have shown that edge filaments in tokamak plasmas can significantly affect the propagation of microwaves across a broad frequency spectrum, resulting in scattering angles of up to $46^\circ$. Parameter scans were carried out for density perturbations comparable in width and amplitude to MAST filaments and the effect on the measured emission was calculated. 3D effects were discovered in the case of an obliquely incident beam.

In general, the problem of EM propagation past wavelength-sized 3D inhomogeneities is not well understood, yet is of importance for both heating and diagnostic applications in the electron cyclotron frequency range for tokamaks, as well as atmospheric physics. To improve this understanding, a new cold-plasma code, EMIT-3D, was written to extend full-wave microwave simulations in magnetized plasmas to 3D, and make comparisons to the existing 2D code IPF-FDMC. This work supports MAST experiments using the SAMI diagnostic to image microwave emission from the plasma edge due to mode conversion from electron Bernstein waves. Significant fluctuations in the SAMI data mean that detailed modelling is required to improve its interpretation.

\end{abstract}

\section{Introduction}

The study of electron cyclotron frequency range EM wave propagation in magnetised plasmas has broad relevance, from ionospheric radio wave studies \cite{budden61} to heating and current drive \cite{prater04} and diagnostics \cite{hartfuss98} in magnetically confined fusion plasmas. While propagation through a homogeneous or slowly-varying monotonic plasma density profile is well understood and can be calculated analytically or using a ray-tracing method, turbulent or otherwise perturbed profiles require detailed numerical modelling using a full-wave method. In order to deal with cases in which the perturbations are 3D in nature, a new 3D full-wave code, EMIT-3D, has been developed to carry out these simulations.

Recently, a diagnostic known as SAMI (Synthetic Aperture Microwave Imaging) has been developed to take advantage of emission due to mode conversion from electron Bernstein waves (EBWs) \cite{shevchenko12}. This is the inverse of a process which has been investigated as a potential means of heating and current drive in overdense fusion plasmas, such as stellarators \cite{laqua97} and spherical tokamaks \cite{urban11}. Installed on the MAST spherical tokamak, the SAMI diagnostic produces high time-resolution ($\sim10\mu$s) 2D images of O-mode emission due to EBWs generated at EC harmonics in the plasma core. These are measured in the 10 - 35 GHz range, with each frequency corresponding to a radial position through the plasma edge \cite{shevchenko11}. However, large fluctuations have been observed in the measured signal during inter-ELM periods; these demand explanation.

A candidate explanation for these fluctuations is the influence of inter-ELM filaments on the mode conversion and propagation of the emission. There is evidence for the presence of filamentary blob structures at the edge of nearly all tokamaks, as well as alternative configurations such as stellarators and reversed-field pinches; they are thought to be responsible for a significant fraction of cross-field transport at the edge of fusion-relevant plasmas \cite{dippolito11}. They have been observed between ELMs on MAST \cite{benayed09}, and since their density range has been measured to be $5 \times 10^{17} - 2 \times 10^{18}$ m$^{-3}$, some filaments will have peak densities close to or above cutoff, particularly at the lower end of the SAMI frequency range. This paper quantifies the effect of typical MAST inter-ELM filaments by varying filament parameters to investigate their interactions with beams.

Previous simulations have been carried out to study RF scattering from cylindrical filaments for ion cyclotron resonance heating (ICRH) and lower hybrid (LH) applications using a full-wave method \cite{myra10}, and spherical blobs for EC and LH waves, using Fokker-Planck and ray-tracing methods \cite{tsironis09}, geometric optics analysis \cite{hizanidis10} and full-wave modelling \cite{ram13}. Since emission can be obliquely incident with a finite width along the filament axis, this scattering problem is in some respects inherently 3D. The new code is therefore used to carry out these simulations in 3D. By comparison with the 2D full-wave code IPF-FDMC \cite{koehn08}, the cases in which 3D treatment is required are determined.

\section{The EMIT-3D code}
\subsection{The FDTD algorithm}

The finite-difference time-domain (FDTD) method is used as the basis for this code. This employs a grid in which field components are staggered both in space and time, allowing the calculation of time-domain solutions to Maxwell's equations via a second-order centred-difference approximation \cite{taflove00}. Fully explicit update equations for electric and magnetic fields are obtained ($\Delta t$ is timestep, $\Delta x=\Delta y=\Delta z$ is spatial step on the Cartesian grid. The ratio $\frac{\Delta t}{\Delta x}$ is constrained by a Courant stability condition. After bar, superscripts are timestep indices, subscripts are spatial coordinates):

\setlength{\mathindent}{0cm}
\begin{equation*}
B_x{}|^{n+\frac{1}{2}}_{i,j+\frac{1}{2},k+\frac{1}{2}} = B_x{}|^{n-\frac{1}{2}}_{i,j+\frac{1}{2},k+\frac{1}{2}} + \frac{\Delta t}{\Delta x} \left[E_y{}|^{n}_{i,j+\frac{1}{2},k+1} - E_y{}|^{n}_{i,j+\frac{1}{2},k} - 
      E_z{}|^{n}_{i,j+1,k+\frac{1}{2}} + E_z{}|^{n}_{i,j,k+\frac{1}{2}}\right] 
\end{equation*}

\vspace{1.5cm}

\begin{eqnarray*}
\setlength{\mathindent}{0pt}
E_x{}|^{n+1}_{i+\frac{1}{2},j,k} = E_x{}|^{n}_{i+\frac{1}{2},j,k} &+& \frac{c^2 \Delta t}{\Delta x} \left[B_y{}|^{n+\frac{1}{2}}_{i+\frac{1}{2},j,k+\frac{1}{2}} - B_y{}|^{n+\frac{1}{2}}_{i+\frac{1}{2},j,k-\frac{1}{2}} - 
B_z{}|^{n+\frac{1}{2}}_{i+\frac{1}{2},j+\frac{1}{2},k} + B_z{}|^{n+\frac{1}{2}}_{i+\frac{1}{2},j-\frac{1}{2},k}\right] \\ 
&-& \frac{\Delta t}{\varepsilon_0}J_x{}|^{n+1}_{i+\frac{1}{2},j,k}
\end{eqnarray*}

with equivalent equations for $y$ and $z$ field components. For propagation through free space, the current density $\mathbf{J} = 0$; however, it is nonzero in a plasma. In order to calculate $\mathbf{J}$, the linearised fluid equation of motion for the electrons must be solved at each timestep: 

\setlength{\mathindent}{2.5cm}
\begin{equation} \label{eq:fluid}
\frac{\partial \mathbf{J}}{\partial t} = \varepsilon_0
\omega_{p e}^2\mathbf{E} - \omega_{c e}\mathbf{J}\times\mathbf{\hat{b}}_0 - \nu\mathbf{J}
\end{equation}

where $\omega_{p e}$ and $\omega_{c e}$ are the local electron plasma frequency and cyclotron frequency, $\mathbf{\hat{b}}_0$ is the unit vector parallel to the background magnetic field and $\nu$ is the electron collision frequency, set to zero on the main grid of the simulation but nonzero in the boundary regions to damp outgoing waves. These parameters modify the value of the Courant limit.

To obtain discretised solutions to this 3D system of equations, Equation~\ref{eq:fluid} is written in matrix form:

\begin{equation} \label{eq:matrix}
\frac{\partial \mathbf{J}}{\partial t} = A \mathbf{J} + \varepsilon_0\omega_{p e}^2 \mathbf{E}
\end{equation}

where 

\begin{equation*}
A = \left( \begin{array}{ccc}
-\nu & -b_z\omega_{ce} & b_y\omega_{ce} \\
b_z\omega_{ce} & -\nu & -b_x\omega_{ce} \\
-b_y\omega_{ce} & b_x\omega_{ce} & -\nu \end{array} \right)
~~\mbox{,}~~
\mathbf{\hat{b}}_0 = (b_x,b_y,b_z)
\end{equation*}

This has the time-invariant solution:

\begin{equation} \label{eq:matrixsoln}
\mathbf{J}(t) = e^{A(t-t_0)}\mathbf{J}(t_0) + \varepsilon_0\omega_{p e}^2 \int_{t_0}^t e^{A(t-\tau)} \mathbf{E}(\tau) \mathrm{d}\tau
\end{equation}

In this discretisation scheme, $\mathbf{J}$ is calculated in phase with $\mathbf{H}$, i.e.\ at $t=(\frac{1}{2},\frac{3}{2},\frac{5}{2},...)\Delta t$, and so a value of $\mathbf{E}$ at $t=(0,1,2,...)\Delta t$ is available. This is used as a midpoint approximation (valid for sufficiently small $\Delta t$) to the value of $\mathbf{E}$ over the interval $[t_0,t]$, allowing $\mathbf{E}$ to be removed from the convolution in Equation~\ref{eq:matrixsoln}.

A simple integration results. After carrying this out and discretising $(\Delta t = t - t_0)$, we obtain:

\begin{equation}\label{eq:disc}
\mathbf{J}{}|^{n+\frac{1}{2}} = e^{A\Delta t}\mathbf{J}{}|^{n-\frac{1}{2}} + \varepsilon_0\omega_{p e}^2 A^{-1}\left(e^{A\Delta t} - I\right)\mathbf{E}{}|^n
\end{equation}

Matrix exponentials must therefore be evaluated; inverse Laplace transforms were used to do this exactly, since $e^{A \Delta t}=\mathcal{L}^{-1} \{ \left(sI - A\right)^{-1} \}^{t=\Delta t}$, although lower-order expansions could also be employed. After some algebraic manipulation, explicit update coefficients for each current density component are obtained, with no storage of old components required after a timestep. As static backgrounds are assumed over the timescale of a single simulation, these update coefficients can be calculated once at the beginning of a simulation and do not require updating.

Data-level parallelisation using MPI is implemented to reduce computation time, which can be quite significant: 3000 timesteps on a $700\times1400\times940$ grid take approximately 7 hours on 80 CPU cores.

\subsection{O and X-mode dispersion relations}

\begin{figure}[h]
\centering
\includegraphics[scale=0.4]{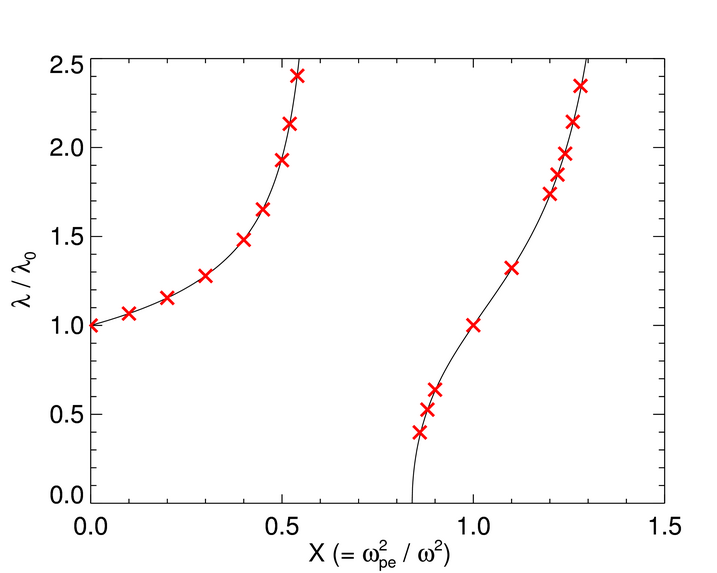}
\caption{X-mode dispersion relation for $Y=|\omega_{ce}|/\omega=0.4$ calculated from averaged wavelength in EMIT-3D output. Wavelength normalised to vacuum wavelength plotted against density normalised to critical density. Red crosses: numerical, Black line: analytic}
\label{fig:xdisp}
\end{figure}

A benchmarking test of the code is its ability to reproduce the analytic dispersion relations for the O- (ordinary) and X- (extraordinary) modes observed in a magnetised plasma. Very close agreement is observed for both; X-mode plotted in Fig.~\ref{fig:xdisp}.

\section{Results}

A filament was modelled as a cylindrical density perturbation with a Gaussian profile:

\begin{equation}
n_e(y,z)=n_{e,\mathrm{peak}} \exp(\frac{-(y-y_0)^2-(z-z_0)^2}{w^2})
\end{equation}

Results were obtained by exciting a linearly polarised (in the $x$-direction) O-mode beam at the $z = 0$ plane, a distance $z = 7 \lambda_0$ from the cylinder axis, where $\lambda_0$ is the vacuum wavelength of the incident beam. After reaching a steady state, the RMS value of the transverse electric field of the scattered beam was calculated at the backplane (also $z = 7 \lambda_0$ from the cylinder), in order to obtain a power distribution.

To estimate the degree of scattering, the mean $\mu_P$ and standard deviation $\sigma_P$ of this power were calculated in both $x$- and $y$-directions as follows (e.g. for $y$):

\begin{equation}
P_{tot} = \sum_{x_0}^{N_x}\sum_{y_0}^{N_y}P(x,y)
\end{equation}

\begin{equation}
\mu_{P,y} = \frac{1}{P_{tot}}\sum_{x_0}^{N_x}\sum_{y_0}^{N_y}[y \cdot P(x,y)]
\end{equation}

\begin{equation}
\sigma_{P,y} = \sqrt{\frac{2}{P_{tot}}\sum_{x_0}^{N_x} \sum_{y_0}^{N_y}[(y-\mu_{P,y})^2 \cdot P(x,y)]}
\end{equation}

where $N_x$, $N_y$ are the total number of gridpoints in the $x$- and $y$-directions respectively. These quantities were compared against their values $\mu_{P,\mathrm{vac}}$, $\sigma_{P,\mathrm{vac}}$ for a beam propagating only through vacuum. The point of maximum emission $y_{\mathrm{max}}$ was also recorded. Parameters were scanned through experimentally relevant values over a number of runs, and the degree of scattering quantified.

\begin{figure}[h!]
\centering
\includegraphics[scale=0.45]{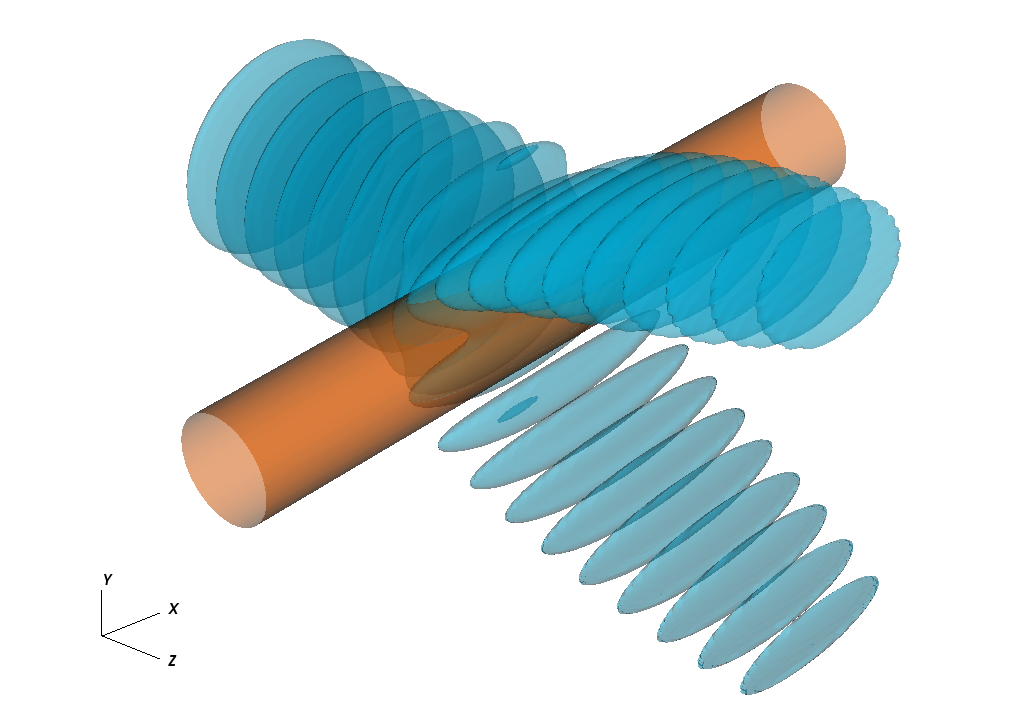}
\caption{3D surface plot of EMIT-3D output showing simulation setup. Blue: $E_x = 0.41A$ surface of beam, orange: $X=0.25$ surface of background density. Incident beam is excited at left-hand plane, scattered by filament at centre of domain, then propagates as two scattered beams towards right-hand plane and is absorbed at boundary}
\label{fig:surf3d}
\end{figure}

\subsection{Scaling to experiment}

To ensure experimental relevance, physical parameters for the investigation were obtained from Ben Ayed \emph{et al.} (2009) \cite{benayed09}. Langmuir probe measurements found that  MAST inter-ELM filaments have a density range of $n_{e} = 5 \times 10^{17} - 2 \times 10^{18}$ m$^{-3}$. For experimentally relevant frequencies of 10 $-$ 35 GHz, this corresponds to a range in the parameter $n_{e,\mathrm{peak}} = 0.03n_{e,\mathrm{crit}} - 1.61n_{e,\mathrm{crit}}$.

Filament widths, defined in \cite{benayed09} as ``twice the difference between the maximum and lowest neighbouring minimum'', were measured in the range $w = 5 - 30$ cm, with a peak at 13 cm and mean at 16 cm. Corresponding Gaussian widths are estimated to be around a quarter of this, giving a range of $1.25 - 7.50$ cm. The frequencies of interest correspond to vacuum wavelengths $\lambda_0 = 0.85 - 3.00$ cm, so a range in Gaussian widths of $w = 0.42 \lambda_0 - 8.82 \lambda_0$ is possible.

\subsection{Filament position}

\begin{figure}[H]
\centering
\includegraphics[scale=0.6]{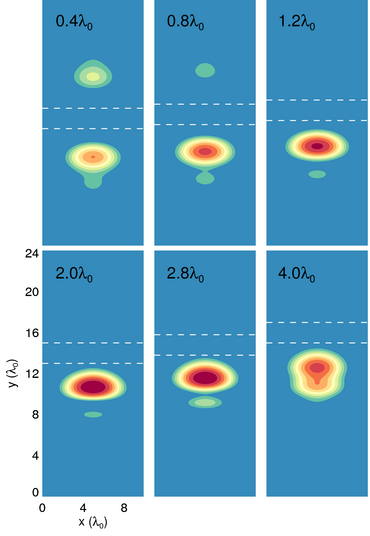}
\caption{Distribution of $\left<E^2\right>$ ($\propto$ power) on backplane for position scan. $y_{\mathrm{fil}}$ shown on each plot. Dashed lines indicate the $\frac{1}{e}$ extent of the filament density projected onto the backplane. Contours on a log scale. Spatial coordinates shown on axes of lower left image}
\label{fig:pos_graphs}
\end{figure}

The position $y_{\mathrm{fil}}$ of the centre of a filament with its axis in the x-direction (normal to beam propagation) was varied relative to the beam axis in the y-direction. The filament's peak electron density was kept constant at $n_{e,\mathrm{peak}} = 0.8 n_{e,\mathrm{crit}}$, and its Gaussian width at $w = 1.0\lambda_0$. Backplane power distributions are plotted in Fig.~\ref{fig:pos_graphs}.

\begin{figure}[H]
\centering
\includegraphics[scale=0.4]{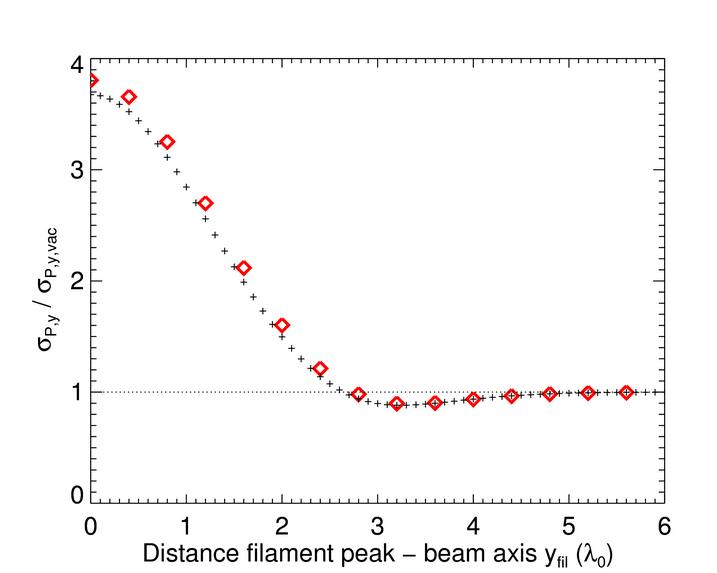}
\caption{Standard deviation $\sigma_{P,y}$ in y-direction for position scan. Dashed line for case without filament. Red diamonds: EMIT-3D, black crosses: IPF-FDMC}
\label{fig:pos_sd}
\end{figure}

Scattering in this case, as could be expected due to the inherently 2D nature of the problem, was only observed in the $y$-direction - in the $x$-direction, the beam profile remained close to the original Gaussian of the source. That is, no scattering out of the plane was observed. The peak $\sigma_{P}$ in the $y$-direction (Fig.~\ref{fig:pos_sd}), observed when the filament and beam axes coincide, is greater than the result for vacuum by a factor of 3. As their separation increases past $2\lambda_0$, power distributions with $\sigma_{P} < \sigma_{P,\mathrm{vac}}$ are observed, suggesting that the filament may be acting to focus the beam, although this is only a small effect.

\begin{figure}[h!]
\centering
\includegraphics[scale=0.4]{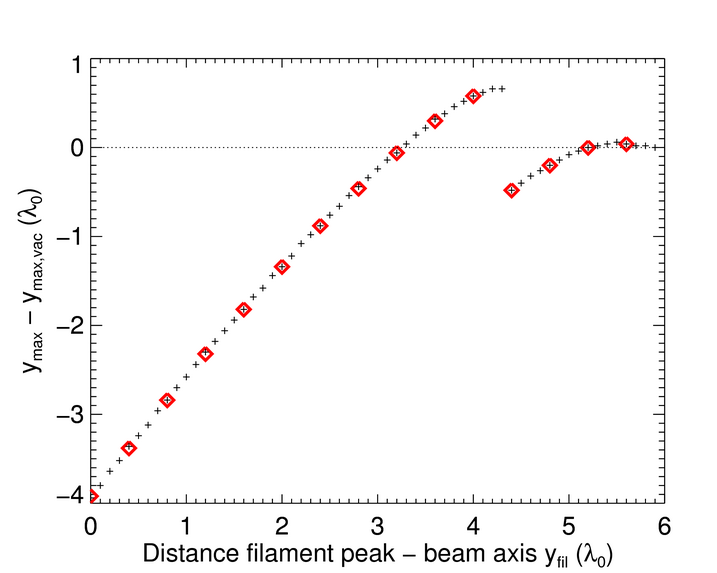}
\caption{Shift of maximum emission point $y_{\mathrm{max}}$ in y-direction for position scan. Dashed line for case without filament. Red diamonds: EMIT-3D, black crosses: IPF-FDMC}
\label{fig:pos_maxem}
\end{figure}

The point of maximum emission $y_{\mathrm{max}}$ (Fig.~\ref{fig:pos_maxem}) was maximally displaced by a filament centred on the beam axis, with two equal maxima located at $y = \pm 3.88\lambda_0$. As the filament moved away, this became a single maximum which returned towards, and then slightly beyond, the centre point at larger separations. A single beam was recovered, but its shape was still influenced by the presence of the filament, with a dual-lobed structure as can be seen in Fig.~\ref{fig:pos_graphs} for $y_{\mathrm{fil}}=4\lambda_0$. 

As the separation increased past $4\lambda_0$, the lower emission lobe became more intense than the upper and hence a discontinuity is seen in the plot of $y_{\mathrm{max}}$. This shows that a filament passing across a beam will in fact cause a series of two perturbations to its maximum, rather than the one which might naively be expected. The greatest value of $y_{\mathrm{max}}$ (with filament aligned to the beam axis) corresponded to a scattering angle of $26^\circ$.

This scan was also carried out in 2D using IPF-FDMC for benchmarking and comparison purposes; 2D results are plotted in black on Figures~\ref{fig:pos_sd} and~\ref{fig:pos_maxem}. Excellent agreement is shown; while it is exact for $y_{\mathrm{max}}$, a small discrepancy in calculated $\sigma_{P}$ is observable for larger scattering angles. However, this is due to necessary differences in the method of analysis between 2D and 3D cases rather than numerical differences.

\subsection{Filament peak density}

\begin{figure}[h!]
\centering
\includegraphics[scale=0.4]{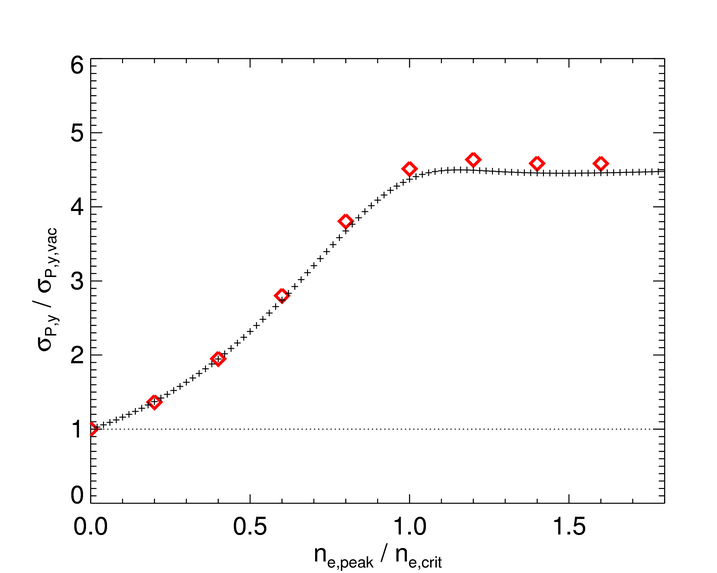}
\caption{Standard deviation $\sigma_{P,y}$ in y-direction for density scan. Red diamonds: EMIT-3D, black crosses: IPF-FDMC}
\label{fig:n_e_sd_y}
\end{figure}

\begin{figure}[h!]
\centering
\includegraphics[scale=0.4]{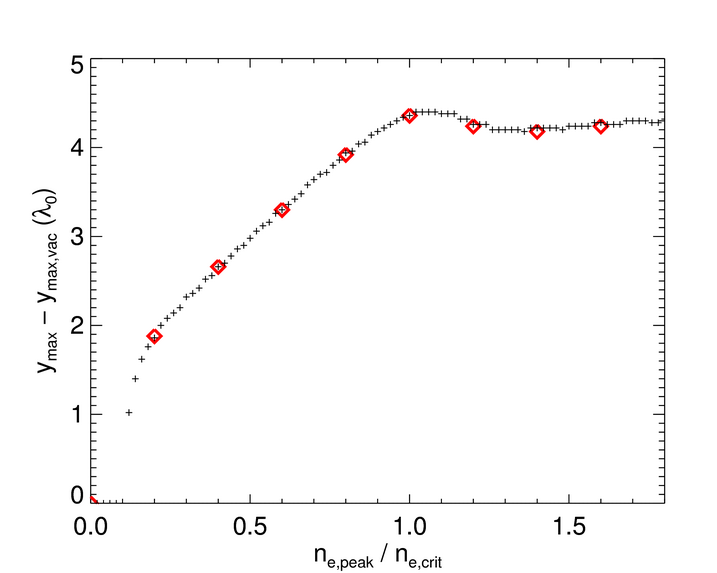}
\caption{Shift of maximum emission point $y_{\mathrm{max}}$ in y-direction for density scan. Red diamonds: EMIT-3D, black crosses: IPF-FDMC}
\label{fig:n_e_maxem_y}
\end{figure}

The filament position was held constant on the beam axis and $n_{e,\mathrm{peak}}$ varied through experimentally relevant values. Again, the scattering observed was 2D in nature. $\sigma_{P}$ (Fig.~\ref{fig:n_e_sd_y}) increased as the density was increased while still below cutoff, reaching a maximum with $n_{e,\mathrm{peak}} = n_{e,\mathrm{crit}}$. As the peak density rose above cutoff, the degree of scattering saturated and remained nearly constant, as did the position of the symmetric emission maxima (Fig.~\ref{fig:n_e_maxem_y}). A maximum scattering angle of $32^\circ$ was not exceeded with increasing $n_{e,\mathrm{peak}}$.

However, the total power reaching the backplane was significantly decreased. Below cutoff, a smaller decrease of total power by $\sim9\%$ was observed due to the beam having scattered beyond the maximum y-extent of the computational domain. As cutoff was reached and exceeded at the filament centre and backscattering began to play a significant role, the total backplane power decreased to $\sim40\%$ of its original value.

Again, excellent agreement of EMIT-3D with IPF-FDMC is seen in Figures~\ref{fig:n_e_sd_y} and~\ref{fig:n_e_maxem_y}.

\begin{figure}[h!]
\centering
\includegraphics[scale=0.6]{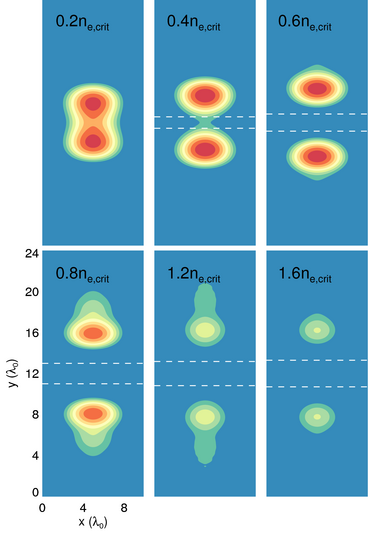}
\caption{As Fig.~\ref{fig:pos_graphs} for filament density scan. $n_{e,\mathrm{peak}}$ shown in terms of $n_{e,\mathrm{crit}}$. Dashed lines at same density contour (not visible for $n_{e,\mathrm{peak}} = 0.2 n_{e,\mathrm{crit}}$ since this contour is not reached. Contours on a log scale. Spatial coordinates shown on axes of lower left image)}
\label{fig:ne_graphs}
\end{figure}

\FloatBarrier

\subsection{Filament width}
The Gaussian width $w$ of the filament was varied from $0 - 2.5\lambda_0$, with fixed peak density $n_{e,\mathrm{peak}} = 0.8 n_{e,\mathrm{crit}}$. For $w < 0.3\lambda_0$, almost no effect on $y_{\mathrm{max}}$ was observed; however, a large increase in $\sigma_y$ was observed for $w > \lambda_0$, with the point of maximum emission also diverging from the mean. A scattering angle of $47^\circ$ was observed at $w = 2.5\lambda_0$, and this continued to increase for wider filaments.

Excellent agreement between codes was again observed.

\begin{figure}[h!]
\centering
\includegraphics[scale=0.6]{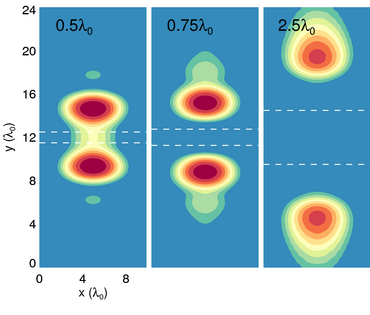}
\caption{As Fig.~\ref{fig:pos_graphs} for width scan. Widths shown in units of $\lambda_0$. Contours on a log scale. Spatial coordinates shown on axes of leftmost image}
\label{fig:w_graphs}
\end{figure}

\begin{figure}[h!]
\centering
\includegraphics[scale=0.4]{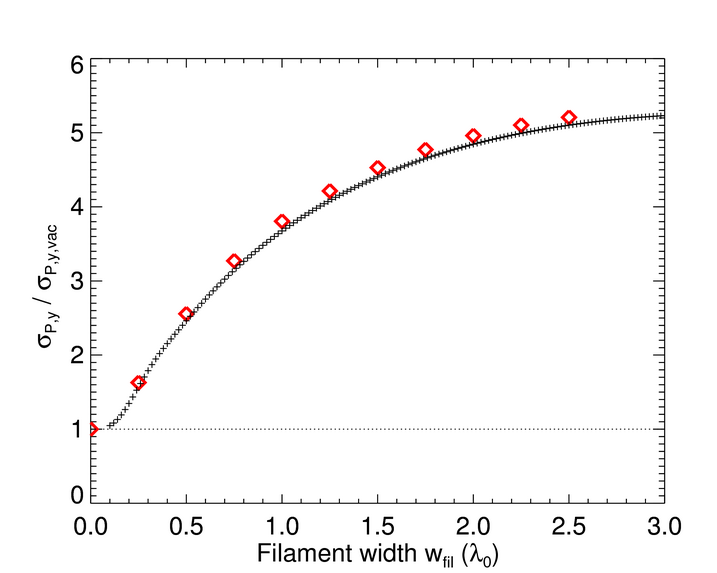}
\caption{Standard deviation $\sigma_{P,y}$ in y-direction for width scan. Red diamonds: EMIT-3D, black crosses: IPF-FDMC}
\label{fig:w_sd_y}
\end{figure}

\begin{figure}
\centering
\includegraphics[scale=0.4]{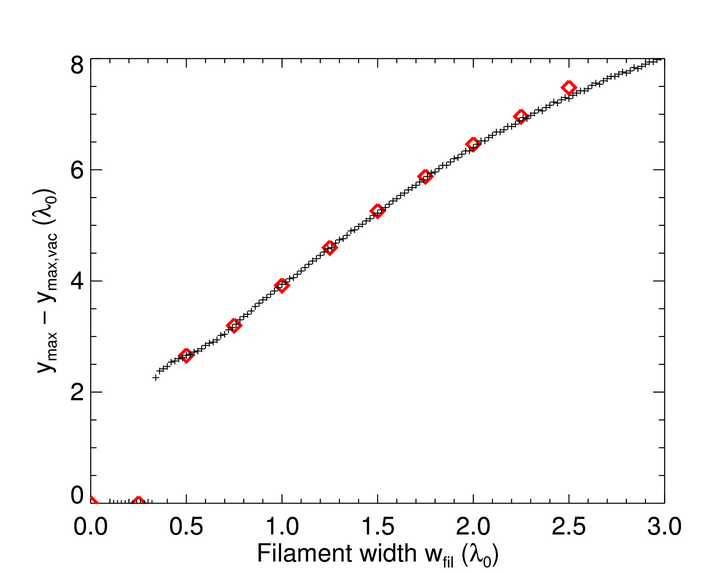}
\caption{Shift of maximum emission point $y_{\mathrm{max}}$ in y-direction for width scan. Red diamonds: EMIT-3D, black crosses: IPF-FDMC}
\label{fig:w_maxem_2}
\end{figure}

\FloatBarrier

\subsection{Filament angle}
\begin{figure}[h!]
\centering
\includegraphics[scale=0.5]{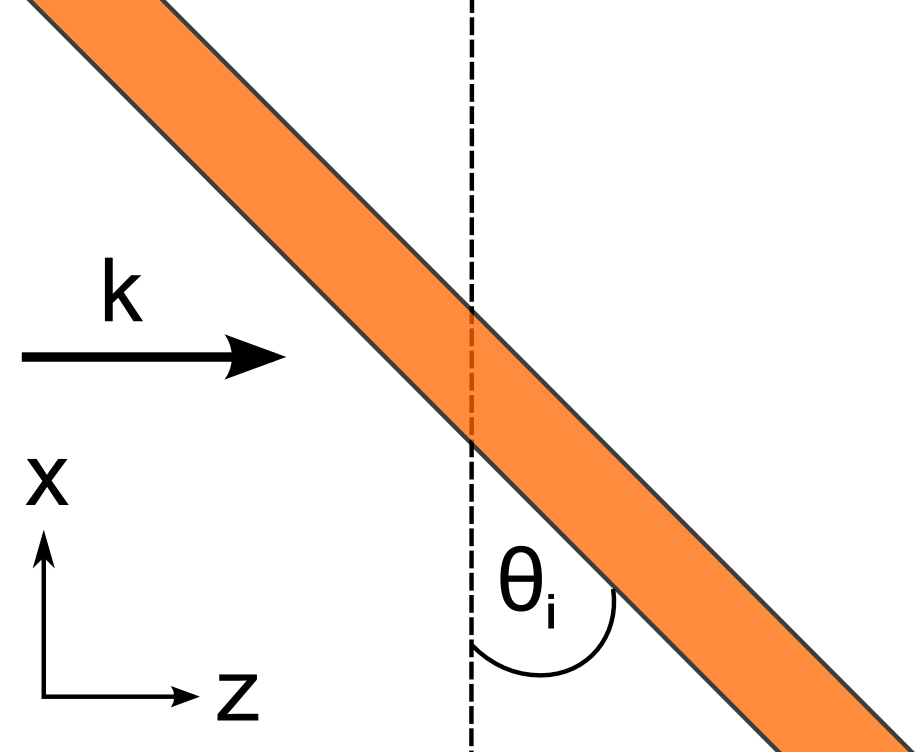}
\caption{Schematic of angular scan in $x$-$z$ plane. Filament tilted at angle $\theta_i$ is shown in orange, $k$ is wavevector of incident beam}
\label{fig:ang_schematic}
\end{figure}

Finally, the incident angle to the normal, $\theta_i$, of the beam on the filament in the $x$-$z$ plane was varied between $0^\circ - 80^\circ$ (Fig.~\ref{fig:ang_schematic}). As for the position scan, filament density and width were held constant at $n_{e,\mathrm{peak}} = 0.8 n_{e,\mathrm{crit}}$ and $w = 1.0\lambda_0$.

As the beam moved away from normal incidence, it encountered an increasingly shallow density gradient and so refractive effects began to dominate over scattering. This resulted in a rotation of the emission patterns, with the two spots coalescing into a single maximum for large angles.

Although the maximum emission point $y_{\mathrm{max}}$ shifted in both $x$- and $y$-directions, the scattering angle only increases slightly ($29^\circ - 31^\circ$). This is an example of a genuinely 3D effect.

\begin{figure}[h]
\centering
\includegraphics[scale=0.6]{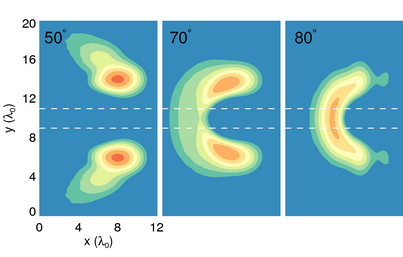}
\caption{As Fig.~\ref{fig:pos_graphs} for angular scan. Incident angles in x-z plane shown. Contours on a log scale. Spatial coordinates shown on axes of leftmost image}
\label{fig:ang_graphs_nb}
\end{figure}

\section{Discussion}

These results show the synergistic effects of two filament parameters, density and width, on the observed scattering angle. While filaments with widths of several times the beam wavelength can cause large deviations at higher frequencies (e.g. a filament of Gaussian width 2.5cm, close to the mean width for MAST, will scatter a 30 GHz beam by $47^\circ$), filament density becomes an important effect at lower frequencies, even for underdense filaments (e.g. a filament of peak density  $1.0\times10^{18}$ m$^{-3}$, again typical for MAST, will scatter a 10 GHz beam by $26^\circ$).

It is therefore shown that filamentary plasma structures can have a significant influence on the propagation of microwaves across a broad frequency spectrum, with consequences for both electron cyclotron diagnostics and heating. Since spherical tokamaks maintain plasma densities comparable to other tokamaks with a lower magnetic field and thus have lower electron cyclotron frequencies, these effects can be particularly important. In particular, inter-ELM filaments on MAST could be a source of fluctuation for the SAMI diagnostic, although further simulation and experimental comparison are required to quantify such a link.

By comparison with IPF-FDMC, the cases in which 2D simulation is justified have been determined. For normal incidence, minimal out-of-plane scattering was observed from EMIT-3D, verifying that 2D simulation, with its reduced computational time, is valid in this case. However, for oblique incidence, which is generally the case for emission diagnostics, 3D effects become important.

Future work using EMIT-3D will seek to quantify the effect of realistic spherical tokamak edge turbulence on both microwave propagation and the mode conversion process. This will be made possible by the integration of turbulent density profiles generated by edge turbulence codes. Furthermore, its 3D geometry will allow the investigation of effects such as magnetic shear \cite{cairns00} and asymmetry of the mode conversion efficiency \cite{gospodchikov06} in regimes where the density scale length $L_{n_e} = \frac{n_0}{|\nabla n_0|} \sim \lambda_0$.

\section*{Acknowledgements}

This work was part-funded by the University of York. It was also part-funded by the RCUK Energy Programme (grant number EP/I501045)] and the European Communities under the contract of Association between EURATOM and CCFE. The views and opinions expressed herein do not necessarily reflect those of the European Commission. To obtain further information on the data and models underlying this paper please contact PublicationsManager$@$ccfe.ac.uk.

\end{document}